\documentstyle[prd,aps,eqsecnum,preprint,tighten,floats]{revtex}
\begin{document}

\title{(1+1)-Dimensional Yang-Mills Theory
Coupled to Adjoint Fermions on the Light Front}

\author{Stephen S. Pinsky} 
\address{Department of Physics, The Ohio State University, Columbus,
OH 43210}

\maketitle

\begin{abstract}
We consider $SU(2)$ Yang-Mills theory in 1+1 dimensions coupled to
massless adjoint fermions.  With all fields in the adjoint
representation the gauge group is actually $SU(2)/Z_2$, which
possesses nontrivial topology.  In particular, there are two distinct
topological sectors and the physical vacuum state has a structure
analogous to a $\theta$ vacuum.  We show how this feature is realized
in light-front quantization, with periodicity conditions in $x^\pm$
used to regulate the infrared and treating the gauge field zero mode
as a dynamical quantity.  We find expressions for the degenerate
vacuum states and construct the analog of the $\theta$ vacuum.  We
then calculate the bilinear condensate in the model.  We argue that
the condensate does not affect the spectrum of the theory, although it
is related to the string tension that characterizes the potential
between fundamental test charges when the dynamical fermions are given
a mass.  We also argue that this result is fundamentally different
from calculations that use periodicity conditions in $x^1$ as an
infrared regulator.

\end{abstract}

\pacs{ }

\section{Introduction}

The unique features of light-front quantization \cite{dir49} make it a
potentially powerful tool for the nonperturbative study of quantum
field theories.  The main advantage of this approach is the apparent
simplicity of the vacuum state.  Indeed, naive kinematical arguments
suggest that the physical vacuum is trivial on the light front.  It is
difficult to imagine that this can really be true, particularly in
light of the important physics associated with the vacuum in QCD.
Thus it is crucial to understand the ways in which vacuum structure
can be manifested in light-front quantization.

This problem has recently been discussed in a variety of contexts.  If
one uses discretization as an infrared regulator (i.e., imposes
periodic or antiperiodic boundary conditions on some finite interval
in $x^-$, the so called discrete light-cone quantization (DLCQ)
approach \cite{pab85}) then any dynamical vacuum structure must
necessarily be connected with the $k^+=0$ Fourier modes of the fields.
Studies of model field theories have shown that certain aspects of
vacuum physics can in fact be reproduced by a careful treatment of the
field zero modes. For example, solutions of the zero mode constraint
equation in $\phi^4_{1+1}$ \cite{may76} exhibit spontaneous symmetry
breaking \cite{hek92a,rob93,bep93,piv94,hip95}.

We shall focus our attention here on $SU(2)$ Yang-Mills theory coupled
to adjoint fermions.  This theory is known to have nontrivial vacuum
structure; as first pointed out in \cite{wit79}, for $SU(N)$ gauge
fields the model has a $Z_N$ topological structure and a physical
vacuum state analogous to a $\theta$ vacuum.  In this theory the
vacuum structure arises because of symmetry considerations and it is
therefore fundamentally different from the $\phi^4_{1+1}$ light-front
theories.  In addition, for $N=2$ there is a nonvanishing bilinear
condensate \cite{smi94}.  This theory is interesting to study for a
variety of additional reasons.  While it differs significantly from
the case of QCD$_{1+1}$ with fundamental matter, where the topology is
trivial and there is a unique vacuum state, it has recently been shown
that the massive spectra of these theories are related \cite{kus95}.
Adjoint fermions are also an important part of the first
(3+1)-dimensional theories where confinement has been proven
\cite{sew94}.

Our goal is to show how this structure is realized in light-front
quantization, using periodicity conditions to regulate the theory. In this
framework it is necessary to introduce degrees of freedom initialized along two
different null planes \cite{mcc88} and to carefully
define singular operator products in a gauge-invariant manner.
Finally, one must pay proper attention to the effects of the boundary
conditions on the choice of gauge.  The most important issues here
concern the proper treatment of the gauge field zero modes.  In
particular, for $SU(2)$ a zero mode of $A^+$ must be retained in the
theory.  This quantity can be treated as a dynamical variable, and
this is the approach we shall take here.  A complementary formalism,
in which this operator is treated as constrained, is discussed
elsewhere \cite{mcr96}.  We shall find two
degenerate vacua and a nonvanishing condensate in the physical ground
state.  Furthermore, the vacuum states can be described completely,
unlike in the equal-time representation.  In this sense the vacuum
structure is simpler on the light front than at equal time.

We will argue that the condensate we find is not related to that
computed at equal time in \cite{les94}.  We also briefly discuss the
implications of our work for the universality of two-dimensional gauge
theories quantized on a light-cone spatial circle \cite{kus95}.  We
will show that the condensate does not affect the massive spectrum of
the theory with massless fermions.  It is however related to the
string tension in the theory with massive adjoint fermions
\cite{grk95}.

\section{$SU(2)$ Gauge Theory Coupled to Adjoint Fermions}

We consider $SU(2)$ gauge theory coupled to adjoint fermions in one
space and one time dimension.  Since all fields transform according to
the adjoint representation, gauge transformations that differ by an
element of the center of the group actually represent the same
transformation and so should be identified.  Thus the gauge group of
the theory is $SU(2)/Z_2$, which has nontrivial topology:
$\Pi_1[SU(2)/Z_2]=Z_2$, so that we expect two topological sectors.
This situation differs from the case when the matter fields are in the
fundamental representation, where the gauge group is $SU(2)$ and the
first homotopy group is trivial \cite{pin94}.

The Lagrangian for the theory is
\begin{equation}
{\cal L} = - {1 \over 2} Tr (F^{\mu \nu} F_{\mu \nu}) + {i \over 2}
Tr(\bar\psi\gamma ^{\mu} \buildrel \leftrightarrow \over
D_{\mu}\psi)\; ,
\end{equation}
where $D_{\mu} = \partial_{\mu} + ig [ A_\mu,\ \ ] $ and $F_{\mu \nu}
= \partial_{\mu}A_{\nu} - \partial_{\nu} A_{\mu} + ig [A_{\mu},
A_{\nu} ]$.  A convenient representation of the gamma matrices is
$\gamma ^0 = \sigma^2$ and $\gamma^1 = i\sigma^1$, where $\sigma^a$
are the Pauli matrices.  With this choice the Fermi field may be taken
to be hermitian.

The matrix representation of the fields makes use of the $SU(2)$
generators $\tau^a=\sigma^a/2$.  It is convenient to introduce a color
helicity, or Cartan, basis, defined by
\begin{equation}
\tau^\pm\equiv {1\over\sqrt{2}}\left(\tau^1\pm i\tau^2\right)\; ,
\end{equation}
with $\tau^3$ unchanged.  Lower helicity indices are defined by
$\tau_\pm = \tau^\mp$.  These satisfy
\begin{eqnarray}
\left[\tau^+, \tau^- \right] &=& \tau^3 \\
\left[\tau^3 , \tau^{\pm} \right] &=& \pm\tau^{\pm}\; .
\end{eqnarray}
In terms of this basis, matrix-valued fields are decomposed as, for
example,
\begin{equation}
A^{\mu} = A _3 ^{\mu} \tau ^3 + A ^{\mu} _+ \tau^+ + A_-^{\mu}\tau^-
\; ,
\end{equation}
where $A^{\mu,\pm}\equiv (A^\mu_1\pm iA^\mu_2)/ \sqrt{2}$ and
$A^{\mu,\pm} = A^\mu_\mp$.  (Note also that
$(A^\mu_+)^\dagger=A^\mu_-$.)  The Fermi field will be written as
\begin{equation}
\Psi _{R/L} = \psi _{R/L} \tau ^3 + \phi _{R/L} \tau ^+ +
\phi^\dagger _{R/L} \tau ^-\; ,
\label{helicity}
\end{equation}
where $\phi_{R/L}\equiv (\Psi^1_{R/L}-i\Psi^2_{R/L})/\sqrt{2}$ and the
labels $R/L$ indicate light-front spinor projections.  Note that under
a gauge transformation the Fermi field transforms according to
\begin{equation}
\Psi _{R/L}^\prime = U \Psi _{R/L} U^{-1}\; ,
\end{equation}
where $U$ is a spacetime-dependent element of $SU(2)$.  The gauge
field transforms in the conventional way.

We shall regulate the theory by requiring that the gauge field
$A^{\mu}$ be periodic and the right-handed Fermi field be antiperiodic
in $x^-$.  The left-handed Fermion $\Psi_L$ is taken to be
antiperiodic in the coordinate $x^+$.  In this ``discretized''
formulation, most of the subtleties have to do with the zero-momentum
Fourier modes of the fields.

The Fock space representation for the Fermi fields is obtained by
Fourier expanding $\Psi_R$ on $x^+=0$ and $\Psi_L$ on $x^-=0$:
\begin{eqnarray}
\psi _R &=& {1 \over 2 ^{1/4} \sqrt {2L}} \sum_n \left(a_n e^{-ik {^+
_n} x^-} + a{_n ^\dagger} e^{ik{_n ^+}x^-} \right) \\
\phi _R &=& {1 \over 2 ^{1/4} \sqrt {2L}} \sum_n \left(b_n e^{-ik {^+
_n} x^-} + d{_n ^\dagger} e^{ik{_n ^+}x^-} \right)\\
\psi _L &=& {1 \over 2 ^{1/4} \sqrt {2L}} \sum_n \left(\alpha_n e^{-ik
{^- _n} x^+} + \alpha{_n ^\dagger} e^{ik{_n ^-}x^+} \right) \\
\phi _L &=& {1 \over 2 ^{1/4} \sqrt {2L}} \sum_n \left(\beta_n e^{-ik
{^- _n} x^+} + \delta{_n ^\dagger} e^{ik{_n ^-}x^+} \right)\; .
\end{eqnarray}
Here the sums run over the positive half-odd integers and $k{^\pm _n}
= n \pi /L$.  The Fourier modes obey the standard anti-commutation
relations
\begin{equation}
\{a{^\dagger _n}, a_m \} = \{ b{_n ^\dagger}, b_m \} = \{ d{_n
^\dagger}, d_m \} = \delta _{n, m}
\label{rhccrs}
\end{equation}
\begin{equation}
\{\alpha{^\dagger _n}, \alpha_m \} = 
\{ \beta{_n ^\dagger}, \beta _m \} = \{
\delta {_n^\dagger},\delta_m\} = \delta _{n, m}\; ,
\label{lhccrs}
\end{equation}
with all mixed anti-commutators vanishing.  These are equivalent to
the canonical anti-commutators
\begin{eqnarray}
\left\{\Psi_R(0,x^-),\Psi_R(0,y^-)\right\}
&=& {1\over\sqrt{2}} \delta (x^--y^-)\\
\left\{\Psi_L(x^+,0),\Psi_L(y^+,0)\right\}
&=& {1\over\sqrt{2}} \delta (x^+-y^+)\; .
\end{eqnarray}
The fermionic Fock space is generated by acting with the various
creation operators on a vacuum state $|0\rangle$.

\section{Current Operators}

The current operators for this theory are
\begin{eqnarray}
J^+ \equiv J^R &=& -{1\over\sqrt 2}[\Psi _{R},\Psi_R]
\label{jplus} \\
\nonumber\\
J^- \equiv J^L &=& -{1\over\sqrt 2}[\Psi _{L},\Psi_L]\; .
\label{jminus}
\end{eqnarray}
To avoid confusion, we shall henceforth always write the currents with
$R$ or $L$ in place of the upper Lorentz index and the color helicity
index either up or down with $J_3=J^3$ and $J_- = J^+$.

These expressions for the currents are ill-defined as they stand since
they contain the product of operators at the same point. This is a
common problem and occurs in the expression for the Poincar\'e
generators as well.  We shall regulate these expressions by point
splitting while maintaining gauge invariance and then take the
splitting to zero after removing the singularities. The singularities
give rise to additional contributions, so called gauge corrections. To
maintain gauge invariance in the split product
one must introduce an eikonal factor.  One can show for example that
\begin{equation}
\left[
e^{ ig \int {_{x} ^{x + \epsilon^\mu}} A \cdot dx}\Psi(x+\epsilon^\mu)
e^{-ig \int {^{x +\epsilon^\mu} _{x} } A \cdot dx},\Psi(x)
\right]
\label{ham}
\end{equation}
transforms canonically in the adjoint representation. In the limit of
small $\epsilon^\mu$ one effectively replaces
\begin{equation}
\psi(x) \rightarrow \psi(x+\epsilon^\mu) +ig  [ A \cdot
\epsilon, \psi( x+\epsilon^\mu) ]
\end{equation}
The singularity in the operator product as $\epsilon \rightarrow0$
picks up the $\epsilon$ in the above expression leaving an additional
contribution.  The splitting must be performed in the $x^-$ direction
for $\Psi_R$.  We find that the current $J^R$ acquires a gauge
correction
\begin{equation}
J^R = \tilde{J^R} - {g \over 2\pi} V\; ,
\label{Ranomaly}
\end{equation}
where $\tilde{J^R}$ is the naive normal-ordered current.  The operator
product in $J^L$ must be split in the $x^+$ direction, and an
analogous calculation gives
\begin{equation}
J^L = \tilde{J^L} - {g \over 2\pi} A \; ,
\label{Lanomaly}
\end{equation}
where $\tilde{J^L}$ is the naive normal-ordered current.

In the helicity basis, $\tilde{J}^R$ takes the form
\begin{eqnarray}
\tilde{J}^R_3 &=& :{1\over\sqrt{2}} \left(
\phi^\dagger_{R} \phi_{R}-\phi_{R} \phi^\dagger_{R} \right): \\
\nonumber\\
\tilde{J}^R_- &=& :{1\over\sqrt{2}} \left(
\psi_{R} \phi^\dagger_{R}-\phi^\dagger_{R} \psi_{R} \right): \\
\nonumber\\
\tilde{J}^R_+ &=& :{1\over\sqrt{2}} \left(
\phi_{R} \psi_{R}-\psi_{R} \phi_{R} \right): \; .
\end{eqnarray}
The corresponding expressions for the components of $\tilde{J}^L$ are
identical, with $R\rightarrow L$.  It is convenient to Fourier expand
these currents and discuss the the properties of their components.  We
write
\begin{eqnarray}
\tilde{J}^{R,a} &=& {1 \over 2L}\sum_{N=-\infty}^{\infty} C^a_N e^{-i\pi N x^-/L}\\
\tilde{J}^{L,a} &=& {1 \over 2L}\sum_{N=-\infty}^{\infty} D^a_N e^{-i\pi N x^+/L}\;
,
\end{eqnarray}
where $a$ is a color index and the sums are over the integers.  It is
well known that the Fourier components satisfy a Kac-Moody algebra
with level one \cite{goo86}.  We shall discuss this explicitly for the
$C_N^a$; an identical set of relations holds for the $D_N^a$, with
appropriate substitutions.

In terms of the Fock operators, we find for $N>0$
\begin{eqnarray}
C_N^3 &=& \sum_{n={1 \over 2}}^\infty b^\dagger_n b_{N+n}
-\sum_{n={1 \over 2}}^\infty d^\dagger_n d_{N+n}
-\sum_{n={1 \over 2}}^{N-{1 \over 2}} b_n d_{N-n}  \nonumber  \\
C_N ^+ &=& \sum_{n={1 \over 2}}^\infty a^\dagger_n d_{N+n}
- \sum_{n={1 \over 2}}^\infty b^\dagger_n a_{N+n}
-\sum_{n={1 \over 2}}^{N-{1 \over 2}} d_n a_{N-n}  \nonumber  \\
C_N ^- &=& \sum_{n={1 \over 2}}^\infty d^\dagger_n a_{N+n}
 - \sum_{n={1 \over 2}}^\infty a^\dagger_n b_{N+n}
-\sum_{n={1 \over 2}} ^{N-{1 \over 2}} a_n b_{N-n}\; .
\end{eqnarray}
The negative frequency modes may be obtained by hermitian conjugation:
\begin{equation}
C^3_{-N} = (C^3_N)^\dagger \qquad C^+_{-N} = (C^-_N)^\dagger
\qquad C^-_{-N} = (C^+_N)^\dagger\; .
\end{equation}
Finally, the $N=0$ terms are just the various color charges in the
right-moving fermions:
\begin{eqnarray}
C^3_0 &=& 
\sum_n (b _n ^\dagger b_n - d_n ^\dagger d_n )\equiv Q_R   \nonumber  \\
C^-_0 &=& 
\sum_n (d _n ^\dagger a_n - a_n ^\dagger b_n )  \nonumber  \\
C^+_0 &=&
\sum_n (a _n ^\dagger d_n - b_n ^\dagger a_n )\; .
\end{eqnarray}
In the Cartan basis, the Kac-Moody algebra takes the form
\begin{eqnarray}
\left [ C^3_N , C^3_M         \right ] &=& N \delta_{N,-M} \\
\left [ C_N^{\pm} , C_M^{\pm} \right ] &=& 0 \\
\left [ C^3_N , C_M^{\pm}     \right ] &=& \pm C_{N+M}^{\pm} \\
\left [ C^+_N , C^-_M         \right ] &=& C^3_{N+M} + N \delta_{N,-M}\; .
\end{eqnarray}
It is straightforward to verify these relations using the fundamental
anticommutators (\ref{rhccrs}) and (\ref{lhccrs}).  The algebra
satisfied by the $D_N^a$ is of course identical.  Note that for
$N=M=0$ the above algebra is the $SU(2)$ algebra of the charges.

\section{Gauge Fixing}

In the present context, the main subtlety arising from discretization
is in fixing the gauge.  It is most convenient in light-front field
theory to choose the light-cone gauge $A^+ \equiv V =0$.  This is not
possible with the boundary conditions we have imposed, however; since
the gauge transformation must be periodic up to an element of the
center of the gauge group (here $Z_2$), we cannot gauge the zero mode
of $V$ to zero \cite{frn81}.  It is permissible to take
$\partial_-V=0$.  In addition, we can make a further global (i.e.,
$x^-$-independent) rotation so that the zero mode of $V$ has only a
color 3 component,
\begin{equation}
V = v (x^+) \tau ^3 \; .
\end{equation}
Note that the gauge transformation required to achieve this is
$x^+$-dependent.  In addition, the gauge field satisfies no particular
boundary conditions in $x^+$.  Thus the necessary transformation can
be expected to conflict with the boundary condition imposed on
$\Psi_L$.  However, we can apply a purely $x^+$-dependent gauge
transformation that restores the antiperiodicity of $\psi_L$ on its
initial-value surface $x^-=0$. 

While one could simultaneously rotate $A^- \equiv A$ so that it has no
color 3 zero mode \cite{kap94,pik95}, we shall not do that here.
Instead we will retain this zero mode, which we call $w(x^+)$, and
determine it in the solution of the equations of motion.

At this stage the only remaining gauge freedom involves certain
``large'' gauge transformations which we shall denote $T^R_N$ and
$T^L_N$, with $N$ any integer:
\begin{eqnarray}
T^R_N &=& \exp\left[ -{ iN \pi \over 2L}  x^-\tau_3\right] \\
T^L_N &=& \exp\left[ { iN \pi \over 2L}  x ^+\tau_3\right]\; .
\end{eqnarray}
This freedom is best studied in terms of the dimensionless variables
$Z_R= g v L/ \pi$ and $Z_L= g w L/ \pi$, which $T^{R/L}_N$ shifts by
$\pm N$:
\begin{eqnarray}
T^R_NZ_R (T^R_N)^{-1} &=& Z_R + N
\label{whattdoes}\\
T^L_NZ_L (T^L_N)^{-1} &=& Z_L - N\; .
\end{eqnarray}
In addition, $T^{R/L}_N$ generates a space-time-dependent phase
rotation on the matter field $\phi _{R/L}$,
\begin{eqnarray}
T^R_N \phi _{R} (T^{R}_N)^{-1} &=& e ^{-iN\pi  x^-/ L} \phi _{R}
\label{tsymm-phir} \\
T^L_N \phi _{L} (T^{L}_N)^{-1} &=& e ^{ iN\pi  x^+/ L} \phi _{L}\; ,
\label{tsymm}
\end{eqnarray}
which however preserves the boundary conditions on $\phi_{R/L}$.  The
combination $T^R_N T^L_N \equiv T_N$ is a gauge freedom and
connects different Gribov copies in the theory \cite{gri78}.  
We can use it to bring $Z_R$ to a ``fundamental modular domain''
(FMD), for example $-1 < Z_R < 0$.  Once this is done all gauge
freedom has been exhausted and the gauge fixing is completed.

The physical degrees of freedom that remain are the Fermi fields
$\Psi_R$ and $\Psi_L$ and the gauge field zero mode $Z_R$, restricted
to the finite interval $-1 <Z_R< 0$.  All other nonvanishing
components of the gauge field will be found to be constrained, as is
usual in light-front field theory.  Because of the finite domain of
$Z_R$, it is convenient to use a Schr\"odinger representation for this
variable.   Thus the states of the theory will
be written in the form
\begin{equation}
|\Phi\rangle = \zeta(Z_R)|{\rm Fock}\rangle\; ,
\end{equation}
where $\zeta(Z_R)$ is a  Schr\"odinger wavefunction on the FMD and $|{\rm
Fock}\rangle$ is a Fock state in the fermionic variables.  There
remains the question of what boundary conditions should be satisfied
by the wavefunction $\zeta$.  A careful analysis of the integration
measure used to define the inner product shows that there is a
Jacobian factor which essentially forces the wavefunction to vanish at
the boundaries of the fundamental domain \cite{letly91}.  Thus we have
\begin{equation}
\zeta(-1)=\zeta(0)=0\; .
\end{equation}

After gauge fixing $T_N$ is no longer a symmetry of the theory, but
there is an important symmetry of the gauge-fixed theory that is
conveniently studied by combining $T_1$ with the so-called Weyl
transformation, denoted by $R$.  Under $R$,
\begin{eqnarray}
RZ_{R/L}R^{-1} &=& -Z_{R/L} \label{rsymm-zr}\\
R \phi_{R/L} R^{-1} &=& \phi^\dagger_{R/L}\; .
\label{rsymm}
\end{eqnarray}
$R$ is also not a symmetry of the gauge-fixed theory, as it takes
$Z_R$ out of the fundamental domain.  However, the combination $T_1R$
{\em is} a symmetry; from Eqs. (\ref{whattdoes}) and (\ref{rsymm-zr})
we have
\begin{eqnarray}
T_1R Z_R R^{-1}T_1^{-1} &=& -Z_R-1 \\
T_1R Z_L R^{-1}T_1^{-1} &=& -Z_L+1\; ,
\end{eqnarray}
so that $T_1R$ maps the FMD $-1<Z_R<0$ onto itself.  In fact, it
represents a reflection of the FMD about its midpoint $Z_R=-1/2$.

The action of $T^{R/L}_1$ and $R$ on the fermion Fock operators can be
determined from Eqs. (\ref{tsymm-phir}), (\ref{tsymm}) and
(\ref{rsymm}).  $T^R_1$ gives rise to a spectral flow for the right
handed fields,
\begin{eqnarray}
T^R_1 b_n T^{-1R}_1 & = & b_{n-1}  \qquad (n > 1/2)\nonumber \\
T^R_1 d_n T^{-1R}_1 & = & d _{n+1} \label{trsymm}\\
T^R_1 b _{1/2} T^{-1R}_1 & = & d {_{1/2} ^\dagger}\;,
\nonumber
\end{eqnarray}
while $T^L_1$ gives rise to spectral flow for the left handed fields,
\begin{eqnarray}
T^L_1 \beta_n T^{-1L}_1 & = & \beta_{n+1} \nonumber \\
T^L_1 \delta_n T^{-1L}_1 & = & \delta _{n-1}\qquad (n >1/2)\\
T^L_1 \delta_{1/2}T^{-1L}_1 & = & \beta{_{1/2}^\dagger}\; .
\nonumber
\end{eqnarray}
$R$ is analogous to charge conjugation:
\begin{eqnarray}
R b_n R^{-1} &=& - d_n \\
R \beta _n R^{-1} &=& - \delta _n\; .
\end{eqnarray}
The $a_n$ and $\alpha_n$ are invariant under both $T^{R/L}_1$ and
$R$. From the behavior of the Fock operator it is straightforward to
deduce the behavior of the elements of the Kac Moody algebra under
$T^R$, $T^L$ and $R$ and show that the algebra is invariant.  We shall
elaborate on the detailed implications of the symmetry $T_1R$ when we
consider the structure of the vacuum state below.

Finally, let us discuss Gauss' law and determine the rest of the
vector potential in terms of the dynamical degrees of freedom.
Gauss's law in matrix form is
\begin{equation}
D_- ^2 A = -g J^R
\label{matrixgauss}
\end{equation}
In the Cartan basis this becomes
\begin{eqnarray}
-\partial {^2 _-} A_3 &=& gJ_3^R
\label{gauss3}\\
\nonumber\\
-(\partial _- \pm igv)^2 A_\pm &=& g J_\pm^R\; .
\label{gausspm}
\end{eqnarray}
This is the reason for introducing the color helicity basis: in the
gauge we have chosen, Gauss' law separates and its individual
components can be solved directly.  Note that because of the gauge
choice, $J^R$ only acquires a gauge correction to its 3 color
component.  All color components of $J^L$ receive a gauge correction.

Eqn. (\ref{gauss3}) can be used to obtain the normal mode part of
$A_3$ on the surface $x^+=0$:
\begin{equation}
A_3 = {gL \over 2\pi^2} \sum_{N \neq 0} { C^3_N \over N^2}
e^{-iN \pi x^-/L}\; .
\end{equation}
The zero mode of $A_3$ is not determined from Eqn. (\ref{gauss3}).  We
shall return to this in a moment, but for now note that since there is
no zero mode on the left hand side of Eqn. (\ref{gauss3}), the zero
mode of the right hand side must also vanish. This gives the relation
\begin{equation}
0=Q_R - Z_R(x^+)\; ,
\label{bigcon}
\end{equation}
Clearly this can not hold as an operator relation.
Eventually this condition must be satisfied by restriction to a
physical subspace.  We shall discuss this in detail below.

Because of the restriction of $Z_R$ to the domain $[-1,0]$ and the
boundary condition on $\zeta(Z_R)$, the covariant derivatives
appearing in Eqn. (\ref{gausspm}) have no zero eigenvalues.  Thus they
may be inverted to give
\begin{equation}
A_\pm = {gL \over 2\pi^2} \sum_{N} {C^\mp_N \over \left (N \mp Z_R
\right )^2} e^{-iN \pi x^-/L}\; .
\end{equation}

\section{Poincar\'e Generators}
The Poincar\'e generators $P^-$ and $P^+$ have contribution from both
the left handed and right handed fermions but $\Psi_R$ is initialized
on $x^+ = 0$ and propagates in $x^+$ while $\Psi_L$ is initialized on
$x^- = 0$ and propagates in $x^-$. These operators must therefore have
contributions from both parts of the initial-value surface.  Thus
\begin{equation}
P^\pm = \int _{-L} ^L dx^- \Theta^{+\pm} + \int _{-L} ^L dx^+
\Theta^{-\pm}\; ,
\end{equation}
where $\Theta^{\mu\nu}$ is the energy momentum tensor.  There are two
common forms for this, the Noether form and the symmetric
gauge-invariant form.  It is instructive to consider both.  The
standard Noether procedure gives
\begin{eqnarray}
\Theta^{+-} &=&-2Tr \left(F^{+-} \partial_+ A^+ \right )
 -Tr\left (F^{+-} F^{+-} \right)
+{i \over \sqrt{2}} Tr \left( \Psi_R \partial^- \Psi_R -\partial^-
\Psi_R \Psi_R \right)\\
\Theta^{-+} &=&-2Tr \left(F^{+-} \partial_- A^- \right )
 -Tr\left (F^{+-} F^{+-} \right)
+{i \over \sqrt{2}} Tr \left( \Psi_L \partial^+ \Psi_L -\partial^+
\Psi_L \Psi_L \right)\\
\Theta^{++} &=& {i \over \sqrt{2}}Tr \left( \Psi_R \partial^+
 \Psi_R -\partial^+ \Psi_R \Psi_R \right)\\
\Theta^{--} &=& {i \over \sqrt{2}}Tr \left( \Psi_L \partial^-
 \Psi_L -\partial^- \Psi_L \Psi_L \right)
-2Tr \left(F^{-+} \partial_+ A^- \right )\;.
\end{eqnarray}
We see that both $\Theta^{-+}$ and $\Theta^{--}$ contain fields that
are initialized on both $x^+ = 0$ and $x^-=0$. What are we to do with
the parts that are initialized on $x^-=0$?  One guess is simply to
throw them away.  Now consider the symmetric gauge-invariant form,
\begin{eqnarray}
\Theta^{+-} &=&Tr \left( F^{+-} F^{+-} \right) \\
\Theta^{-+} &=&Tr \left( F^{+-} F^{+-} \right) \\
\Theta^{++} &=& {i \over \sqrt{2}}Tr \left( \Psi_R D^+ \Psi_R
-D^+ \Psi_R \Psi_R \right)\\
\Theta^{--} &=& {i \over \sqrt{2}}Tr \left( \Psi_L D^- \Psi_L
-D^- \Psi_L \Psi_L \right)\;.
\end{eqnarray}
Again $\Theta^{--}$ and $\Theta^{+-}$ are ambiguous and if we use our
rule of simply dropping terms then there is no reason to believe that
these two forms are equivalent or that either is correct.  Given this
unpleasant situation we chose to use the symmetric gauge-invariant
form and and apply the rule of dropping terms that are not initialized
on the surface that they are to be integrated on. In the end we will
justify this decision by checking that these generators result in the
correct equation of motion.

First let us construct $P^-$. The contribution from the $x^+ = 0$
surface is the standard form one expects in theories of $QCD_{1+1}$
coupled to matter and includes the contribution from the zero mode of $A^+$. 
In the contribution from the
$x^-=0$ surface we will drop the terms containing $A$ in the covariant
derivative. This leaves us with the same contribution that we would obtain from the
Noether current applying the rule and this term give the momentum of the left
movers which propagate in the $x^-$ direction.  We thus obtain
\begin{equation}
P^- = -g^2 \int ^L _{-L} dx^- Tr \left( J^+ {1\over D_-^2 } J^+ \right)
+  L(\partial_+ v)^2 +
+ {i \over \sqrt{2}}\int _{-L} ^L dx^+ Tr \left( \Psi_L \partial_+ \Psi_L
+ \partial_+ \Psi_L \Psi_L \right)\; .
\label{basicham}
\end{equation}
Let us consider the left and right handed parts of this expression
separately for a moment. The operator products appearing in the left-handed part
are singular and require regulariztion and renormalization.  We have done this in two ways,
using the gauge-invariant point-splitting discussed earlier and also a
$\zeta$-function regularization.  Both procedures give the same
result:
\begin{equation}
P^-_{lh} = {\pi \over L } {\sum_n}' n (\alpha{_n ^\dagger} \alpha_n +
\beta{_n^\dagger} \beta_n + \delta{_n ^\dagger}\delta_n ) -
{\pi \over L} Z_L Q_L +{\pi \over 2L} Z^2_L.
\label{hamleft}
\end{equation}
We will have more to say about these extra term as we
further develop this theory.

Now consider the right handed contribution. This takes its most
elegant form by expressing the $J^R$s in term of the $C^N$s. We find
\begin{equation}
P^-_{rh} = {g^2 L \over 4\pi^2} \left[
\sum_{N \neq 0}{C^3_N C^3_{-N} \over N^2} +
\sum_N \left [
{C^+_N C^-_{-N} \over (Z_R+N)^2} + {C^-_{N} C^+_{-N} \over (Z_R-N)^2} \right ]
+\Pi^2_R \right]
\label{hamright}
\end{equation}
where $\Pi_R=(2 \pi/g)\partial_+ v$ is the momentum
conjugate to $Z_R$, so that $[Z_R,\Pi_R]=i$.

Turning now to the operator $P^+$ we see that it only gets contributions
from the $x^+=0$ surface when using the symmetric gauge invariant
form. Thus the longitunial momentum operator is given by
\begin{equation}
P^+ = {g \over \sqrt{2}}\int _L ^L dx^- Tr \left(
\Psi_R  D_+ \Psi_R + D_+ \Psi_R \Psi_R \right)
\label{p+}
\end{equation}

These operator products  are again singular and require point splitting. We find,
\begin{equation}
P^+ = {\pi \over L } {\sum_n} n (a{_n ^\dagger} a_n + b{_n^\dagger}
b_n + d{_n ^\dagger}d_n ) - {\pi \over L} Z_R Q_R + {\pi
\over 2L} Z_R^2\; .
\label{qz}
\end{equation}

One can explicitly show that this expression is $T_1$ and $R$
invariant. We will modifiy this express further below went we discuss the
equation of motion and the conditions that define the physical subspace.

It is interesting to note that for the Noether form there would
appear to be a very substantial contribution from the $x^-=0$
surface. The application of our rule for that term is somewhat unclear
here because the term contains a derivative perpendicular to the
$x^+=0$ surface. Using the equation of motion for $\Psi_L$ we find that the
Noether form for $P^+$ is
\begin{equation}
P^+_N= {\pi \over L} \sum_n n(a{_n ^\dagger} a_n + b{_n^\dagger}
b_n + d{_n ^\dagger}d_n ) + {\pi \over L} \int^L_{-L} dx^+ Z_R(x^+) 
\tilde {J}^L_3(x^+).
\label{pplusn}
\end{equation}
This form is of course not $T$ invariant. Furthermore $Z_R(x^+)$ is
initialized on $x^+ =0$ and not on $x^-=0$ where the left handed contribution
to $P^+$ is calculated. Formally one can envisage solving for $Z_R(x^+)$ in
terms of the operator initialized on $x^+ =0$  and then using it in the above
expression. This of course raises the speculative nature of this form of $P^+$
to an even higher level, and we only mention this form of $P^+$ because it generates
an interesting result in a later section.

\section{Vacuum States of the Theory}
The Fock state containing no particles will be called $\vert
V_0\rangle$.  It is one of a set of states that are related to one
another by $T_1$ transformations, and which will be denoted
$|V_M\rangle$, where $M$ is any integer.  These are defined by
\begin{equation}
|V_M\rangle\equiv (T_1)^M|V_0\rangle\; ,
\end{equation}
where $(T_1)^{-1}=T_{-1}$.  It is straightforward to determine the
particle content of the $|V_M\rangle$.  Consider, for example, the
$T^R_1$ transform of
\begin{equation}
b {^\dagger _{1/2}} b_{1/2} \vert V_0 \rangle = 0\; ,
\end{equation}
which is
\begin{equation}
T^R_1 b{^\dagger _ {1/2}} T^{R-1}_1 T^R_1 b_{1/2} T^{R-1}_1 T^R_1 \vert V_0
\rangle = 0\; .
\end{equation}
Using Eqn. (\ref{trsymm}) we have
\begin{equation}
d_{1/2} d{^\dagger _{1/2}} \vert V_1 \rangle = 0
\end{equation}
which implies
\begin{equation}
d{^\dagger _{1/2}} d_{1/2} \vert V_1 \rangle = \vert V_1 \rangle
\end{equation}
and therefore $\vert V_1 \rangle$ will have one $d_{1/2}$ background
particle. Similarly one could apply $T^L_1$ to
\begin{equation}
\delta {^\dagger _{1/2}} \delta_{1/2} \vert V_0 \rangle = 0\; ,
\end{equation}
and one would conclude that $\vert V_1 \rangle$ also contains one
$\beta_{1/2}$ particle. One can show that $\vert V_1 \rangle$ has no
other content. Thus one finds that
\begin{equation}
\vert V_1 \rangle = d^\dagger_{1/2} \beta^\dagger _{1/2} \vert 0 \rangle.
\end{equation}
We will focus here on the Fock states $\vert V_0 \rangle$ and $\vert V_1
\rangle$ since we will find that the  degenerate vacuum state will be constructed
from them.  The combination $T_1R$, which is a symmetry of this theory,
interchanges the Fock states
$\vert V_0
\rangle$ and
$\vert V_1 \rangle$ up to a phase,
\begin{eqnarray}
T_1R \vert V_0 \rangle & = & (phase) \vert V_1 \rangle \\ \nonumber
T_1R \vert V_1 \rangle & = & (phase) \vert V_0 \rangle \; .
\end{eqnarray}
It also might be possible to construct Bloch wave type states that sums over all
domains as discussed in ref. \cite{las94} that are $T_1R$ invariant.

All of our states, as we noted previously, will be constructed from a
Schr\"odinger wave function $\zeta_0(Z_R)$ and a Fock state. This wave function is
the ground state of the Schr\"odinger equation obtained by projecting out the
empty Fock state sector of the Hamiltonian \cite{pim96}. 
 
We find that under $T_1R$ symmetry
\begin{equation}
T_1R \zeta_0(Z_R) = (phase) \zeta_0(-Z_R-1)
\end{equation}
We chose the arbitrary phases such that,
\begin{equation}
T_1R  \zeta_0(Z_R) \vert V_0 \rangle = e^{i\theta} \zeta_0(-Z_R-1) \vert V_1
\rangle
\end{equation}
Then we can now construct the ``$\Theta $ vacuum " for this theory that is invariant
under the $T_1R$ symmetry,
\begin{equation}
\vert \Omega \rangle = { 1 \over \sqrt{2}} \left [ \zeta_0(Z_R) \vert V_0 \rangle
+ e^{i\theta} \zeta_0(-Z_R-1) \vert V_1\rangle \right ]
\label{omega}
\end{equation}
The state we have 
constructed in Eqn. (\ref{omega}) is an eigenstate:
\begin{equation}
T_1R \vert \Omega \rangle= \vert \Omega \rangle .
\end{equation}
It is typically necessary to build the theory on such a vacuum state in order to satisfy
the requirements of cluster decomposition as well.

It is straight forward to show that $\mid \Omega \rangle$ satisfies all the
physical subspace conitions.  At $x^+=0$ one can to show that
\begin{equation}
\langle \Omega \vert Q_R-Z_R\vert \Omega \rangle =0\; ,
\label{g3zmme}
\end{equation}
independent of the precise form of the wave function
$\zeta_0(Z_R)$---the result only depends on the fact that the
wavefunctions multiplying $|V_0\rangle$ and $|V_1\rangle$ are related
in the way shown in (\ref{omega}).  In addition, we must show that the
matrix elements of all $x^+$ derivatives of $Q_R-Z_R$ at $x^+=0$
(i.e., commutators of $Q_R-Z_R$ with $P^-$) vanish.  This follows
trivially, however, since $|\Omega\rangle$ is an eigenstate of $P^-$.
Thus the condition (\ref{g3zmme}) is satisfied for the vacuum state
(\ref{omega}).

The question of whether there are excited states that satisfy
(\ref{g3zmme}) is more difficult. The energy eigenvalues of the system are
proportional to $2L$, as one would expect, since they correspond to fluctuation of
the flux around the entire spatial volume. We will make a large $L$ approximation and
only retain the ground state wave function $\zeta_0(Z_R)$

Then the  physical states are constructed by applying right-handed, color singlet
Fock operators to the ground state\cite{dek93}. With this picture of physical
states we can now consider the remaining anomaly condition Eq.(
\ref{bigcon}).  This condition on physical states (PS) is
\begin{equation}
0= \langle P \; S\vert Q_R-Z_R(x^+) \vert P \; S\rangle \; .
\label{weakcon}
\end{equation}
To satisfy Eq.(\ref{weakcon}) it is sufficient that the anomaly condition Eq.(
\ref{bigcon})  vanish between vacuum states, which we have already shown.

For this condition to hold at all times it is necessary that the matrix
elements between physical states of all time derivatives of this equation also vanish.
It is again sufficient that the first time derivative of $Z_R$ vanish between
vacuum states, 
\begin{equation}
0= \langle \Omega \vert \Pi_R\vert \Omega \rangle \; .
\end{equation}
This condition is again satisfied independent of the form of the wave function
$\zeta_0(Z_R)$. The next condition one gets by taking the commutator of $\Pi_R$
with
$P^-$. Since there are terms in $P^-$ that depend on $Z_R$ this will produce the
following operator.
\begin{equation}
\sum_N {\left \{C_N^+, C_{-N}^-\right \} \over \left (Z_R+N \right )^3}.
\label{difpi}
\end{equation}
The matrix elements of this operator between physical states are not zero and
therefore the condition Eq.(\ref{weakcon}) in not entirely stable. This is to be
expected because  of the large-$L$ approximation that we made. If our physical
subspace formed a complete set of energy eigenstates then the vanishing of all
matrix elements of
$\Pi_R$ would assure the vanishing of all matrix element of the commutators of
$P^-$ with $\Pi_R$. However, since $P^-$ is $Z_R$-dependent it will generate
transitions from the ground state
$\zeta_0(Z_R)$ to the excited flux states with energies proportional to $2L$ which
we have excluded. Neglecting these states necessarily destroys the stability. 
Since all the states that we neglected are high energy states which decouple in
the large $L$ limit, the the instability presumably disappears in the large $L$
limit.  It is interesting to also remark that in the absence of matter the
operator
$\Pi_R$ is constant \cite{kap94} and this problem does not arise.

\section{Equations of Motion}

We have obtained expressions for $P^+$ and $P^-$ in the previous
section but we can not be sure that they are correct for this theory
until we show that they generate the Dirac equation and Ampere's law
for this theory.

In matrix form, the Dirac equations of motion for the theory are
\begin{eqnarray}
D_+ \Psi _R &=& \partial _+ \Psi _R + ig [A, \Psi _R ] = 0
\label{matrixpsir}\\
\nonumber\\
D_-\Psi _L &=& \partial _- \Psi _L + ig [ V, \Psi _L ] = 0
\label{matrixpsil}
\end{eqnarray}
It is straightforward to resolve these into color components, using
the helicity basis described above, and implementing the gauge we have
chosen.  Eqns. (\ref{matrixpsir}) and (\ref{matrixpsil}) become
\begin{eqnarray}
\partial_-\psi_L &=& 0\\
\nonumber\\
\partial_-\phi_L &=& -igv\phi_L\\
\nonumber\\
\partial_+\psi_R &=& ig\left[A_-\phi_R - A_+\phi_R^\dagger\right]\\
\nonumber\\
\partial_+\phi_R &=& ig\left[A_+\psi_R - A_3\phi_R\right]\; .
\end{eqnarray}
The first two equations evolve $\Psi_L$ in $x^-$, off of its initial-value
surface, while the second pair determine the $x^+$-evolution of
$\Psi_R$.

Ampere's law,
\begin{eqnarray}
\nonumber\\
D_+D_- A &=& gJ^-\; ,
\label{matrixampere}
\end{eqnarray}
requires particularly careful consideration in this theory since it
explicitly connects left- and right-handed quantities.  It decomposes
into
\begin{eqnarray}
\partial_+ \partial_- A_3 - \partial_+^2v
+ ig \left[ A_+(\partial_- - igv)A_-
        -A_-(\partial_- + igv)A_+ \right] &=& gJ^-_3
\label{ampere3}\\
\nonumber\\
\partial_+ \left[ (\partial_- \pm igv)A_\pm \right]
\pm ig A_\pm\partial_+v
\pm ig\left[ A_3(\partial_-\pm igv)A_\pm
        -A_\pm \partial_-A_3 \right] &=& gJ^-_\pm\; .
\label{amperepm}
\end{eqnarray}

Checking these is a straightforward exercise in commuting fields with
$P^\pm$ and comparing the results with the corresponding equations of
motion.  We simply summarize the results here.

It turns out that the equation of motion for $\psi_R$ is satisfied if
\begin{equation}
Z_L = -Q_R\; ,
\label{cl}
\end{equation}
which we shall take to be a strong (operator) equality.  This
determines the zero mode of $A_3$, which we were not able to fix using
Gauss' law.  In addition, the color 3 component of Ampere's law,
Eqn. (\ref{ampere3}), leads to
\begin{equation}
Z_L = Q_L\; ,
\end{equation}
which, when combined with Eqn. (\ref{cl}), gives
\begin{equation}
Q_R + Q_L = 0\; .
\label{neutrality}
\end{equation}
We shall impose this as an eigenvalue condition on states.  Ampere's
law also leads to the conditions
\begin{equation}
D_N^3 = D_N^\pm = 0\; ,
\end{equation}
which must be realized in matrix elements between states.  We shall
therefore require physical states to satisfy
\begin{equation}
D_N^3 |\Phi\rangle= D_N^\pm |\Phi\rangle=0\qquad (N>0)\; .
\label{dn3+-}
\end{equation}
Finally, Ampere's law requires the condition (\ref{bigcon}) which we discussed in
the previous section.

We can now modify $P^\pm$ so that the kinematic Heisenberg
equations are satisfied by replacing troublesome
terms with operators to which they are weakly equivalent.  In Eqn. (\ref{hamleft})
we substitute the operator relation (\ref{cl}) and replace
$Q_L
\rightarrow -Q_R$, as suggested by Eqn. (\ref{neutrality}).  The result is
\begin{equation}
P^-_{lh} = {\pi \over L } {\sum_n} n (\alpha{_n ^\dagger} \alpha_n +
\beta{_n^\dagger} \beta_n + \delta{_n ^\dagger}\delta_n ) - {\pi \over
2L} Q_R^2\; ,
\label{finalhamleft}
\end{equation}
which is consistent with the kinematical Heisenberg relation for
$\phi_L$.  Similarly, in Eqn. (\ref{qz}) we can replace $Z_R
\rightarrow -Q_L$ and $Q_R \rightarrow - Q_L$ to obtain
\begin{equation}
P^+ = {\pi \over L } {\sum_n} n (a{_n ^\dagger} a_n + b{_n^\dagger}
b_n + d{_n ^\dagger}d_n ) - {\pi \over2 L} Q_L^2\; .
\label{qzfinal}
\end{equation}
This expression is consistent with all relevant Heisenberg equations.
Note that the replacement $Z_R\rightarrow -Q_L$, which might have been
expected to be inconsistent with the Dirac equation for $\phi_L$, is
in fact consistent in the subspace defined by (\ref{bigcon}).  Thus
this modification of $P^+$ does not lead to any {\em new} conditions
on states.  

The final result is that with $P^+$ given by Eqn. (\ref{qzfinal}) and
$P^-$ given by the sum of (\ref{finalhamleft}) and (\ref{hamright}),
all Heisenberg equations are correctly obtained in the subspace
defined by (\ref{neutrality}), (\ref{dn3+-}), and (\ref{bigcon}).

Let us now discuss the axial anomaly in this theory,
\begin{equation}
\partial _{\mu} J^{\mu,3}_5 = {g \over 2 \pi} \epsilon ^{\mu \nu}
F_{\mu \nu}^3 \; ,
\end{equation}
where $J^{\mu}_5={-1\over\sqrt{2}}[\Psi,\gamma^\mu\gamma^5\Psi]$ and is
related to the vector current through $J^{\mu}_5 = (J^{R}, -J^L)$. It
can be shown that covariant derivatives reduce to partial derivatives
for $J^{\mu,3}$ so the conservation equations below take the Abelian
from for the matter currents.  We can calculate $\partial_+
\tilde{J}{^{R} _3} $ from
\begin{equation}
i[P ^- , \tilde {J}^{R}_3] = \partial _+ \tilde {J}^{R}_3 (x)\; .
\end{equation}
Using
\begin{equation}
[\tilde {J} {^{R} _3} (0,x^-) , \tilde {J} {^{R} _3} (0,y^-)] = 
{i \over 2 \pi} \delta'(x^- - y^-)
\end{equation}
and we find
\begin{equation}
\partial_+ \tilde {J} {^{R} _3} ={g \over 2 \pi} \partial _- A _3\; .
\label{jr}
\end{equation}
Combining this with the gauge correction for $J^R_3$ we find;
\begin{equation}
\partial _+{J} {^{R} _3} ={g \over 2 \pi}( \partial _- A _3 - \partial_+ v)\;.
\end{equation}

Now $\tilde{J}^L _3 $ is explicitly independent of $x ^-$ by the
equation of motion for $\Psi_L$.  However this is deceptive and we see that
if we calculate $ \partial_- \tilde {J} ^L_3 $ from $i[P^+, \tilde {J}^L_3]$
and use  Eq.(\ref{pplusn}) and 
\begin{equation}
[\tilde {J} {^{L} _3} (x^+,0) , \tilde {J} {^{L} _3} (y^+,0)] = 
{i \over 2 \pi} \delta'(x^+ - y^+),
\end{equation}
then we find,
\begin{equation}
\partial_- \tilde {J}{^L_3 }= {g \over 2 \pi} \partial_+ v \approx
0\; ,
\label{jl}
\end{equation}
If we had used the correct  forms of $P^+$,Eq.(\ref{qzfinal}), we would not find
this contribution. On the other hand $\partial_+v$ is $\Pi_R$ and we have already
discussed the fact that all matrix elements of it between physical states
vanish, and this is the reason for the apparent asymmetry between Eq.(\ref{jl})
and Eq.(\ref{jr}). If we retain this term and  combine it with the gauge
correction to
$J ^L _3$ the complete result for $\partial _- J^L _3$ is;
\begin{equation}
\partial _- J{_3 ^L} = -{g \over 2 \pi} (\partial _- A_3- \partial_+v)  \; .
\end{equation}
Therefore
\begin{equation}
\partial _{\mu} J {_ 3 ^{\mu}} = \partial _+ J {_3 ^{R} } +\partial _-
J{_3 ^L} = 0
\end{equation}
and
\begin{equation}
\partial _{\mu} J {^{5, \mu} _3} = \partial _+ J {_3 ^{R}} -
\partial_- J {_3^L} = {g \over \pi} (\partial _- A_3- \partial_+ v) =
 {g \over 2 \pi} \epsilon ^{\mu \nu} F_{\mu \nu}^3 \; ,
\end{equation}
as expected.

\section{The Condensate}

It is generally accepted that $QCD$ in 1+1 dimensions coupled to
adjoint fermions develops a condensate $\Sigma \equiv\langle\Omega|
{\bar\Psi}\Psi |\Omega\rangle$.  Thus far $\Sigma$ has been calculated
in the large-$N_c$ limit \cite{koz95} and in the small-volume limit
for $SU(2)$ in Ref. \cite{les94}.  Previously we calculated a
condensate in the chiral theory with only right-handed fermions
\cite{pir96}. In that calculation it was the field itself that had a
condensate and the result was fundamentally different from what we are
considering here.

The two vacuum states $\zeta_0(Z_R)\vert V_0 \rangle$ and $ \zeta_0(-Z_R-1)\vert
V_1 \rangle$ are exact ground states.  We have a spectral flow associated with the
right- and left-handed operators, and thus the two physical spaces in
the fundamental domain differ by one right-handed and one left-handed
fermion.  This effectively block diagonalizes $P^-$ into two sectors,
one of which is built on a state with no background particles and one
of which is built on a state with background particles.  The two
sectors communicate through the condensate.  Physical states must
respect the $T_1R$ symmetry and must therefore be built on the
``$\theta$ vacuum" $|\Omega\rangle$, which is a linear combination of
these two vacuum states.

To calculate the condensate $\Sigma$ let us consider $ Tr(\bar
{\Psi}(0) \Psi(0))$ and retain only the terms which can contribute to
the vacuum expectation value:
\begin{equation}
Tr(\bar {\Psi}(0) \Psi(0)) = - {i \over 2L \sqrt{2}} \left [
d_{1/2}^\dagger \beta_{1/2}^\dagger +d_{1/2}\beta_{1/2} \right]+\dots
\;.
\end{equation}
We chose to evaluate $\Sigma$ at the point $(0,0)$ because it is the point where
the two initial value surfaces intersect. Because of the approximation we have
made we are unable to preform the calculation away from this point. Taking the
expectation value in
$\vert
\Omega
\rangle$ we find that only the cross terms contribute.  We thus obtain for the
vacuum expectation value of ${\bar \Psi} \Psi$ from Eqn. (\ref{omega}):
\begin{equation}
\Sigma = 
-{\sin\theta \over 2L \sqrt{2} } \int_{-1}^0 \zeta_0(Z_R)\zeta_0(-Z_R-1)
dZ_R.
\label{condensate}
\end{equation}
Eqn. (\ref{omega}) is an exact expression for the vacuum,
Eqn. (\ref{condensate}) is an exact expression for $\Sigma$. We see
that this expression behaves like $1/L$. This is a common result for
discrete light cone calculations and is found even in the Schwinger
model where the exact result is known not to have this behavior.  We
also note that the only other calculation \cite{les94} of this
quantity, done in different gauge, in a small volume approximation and
discretized and quantized at equal time, finds this $1/L$ behavior.

In \cite{pim96} the explicit form of the Hamiltonian for ground state wave function
is given and solved numerically. Using this form of  numerical solution the value
for the condensate is calculated.

We have not addressed directly the question of whether the condensate
affects the spectrum of this theory but clearly it raises a number of
interesting questions.
It is interesting to consider this in light of work of Kutasov and
Schwimmer \cite{kus95}. They claim that there are classes of theories
that have the same massive spectrum.  The simplified version of the
their argument goes as follows: In light-front quantization of
$QCD_{1+1}$ coupled to massless fermions, the left- and right-moving
fermions decouple and the right-movers alone generate the massive
spectrum.  The role of left-moving particles is simply to cancel the
anomaly and balance the total charge of states.  Thus one could in
general choose any left-handed particles that accomplish this.  For
example, in the theory considered here (one adjoint fermion) we could
just as well take for the left-handed fields two flavors of
fundamental fermions.  Then one argues, based on Lorentz and gauge
invariance, that one could study this new theory in the gauge $A^-=0$
and using $x^-=0$ as the principal quantization surface.  In this case
the roles of the left- and right-moving particles are reversed, and
the spectrum is now generated by the left-handed particles.  The
massive spectra (as well as scattering amplitudes for massive states)
in these models should therefore be the same.  For example,
$QCD_{1+1}$ coupled to two flavors of fundamental fermions should have
the same massive spectrum as a single flavor of adjoint fermions.
However the theory with fundamental fermions will not have a
condensate.

One of the conditions for this universality is the decoupling of the
left- and right-handed fields.   This follows because in the basis
built on $\vert V_0 \rangle$ and $\vert V_{-1} \rangle$, $P^-$ is 
block diagonal.  Furthermore, using the $T_1R$ symmetry one can show
that the blocks are identical.  Therefore the condensate does not
affect the massive spectrum of the theory.

Kutusov and Schwimmer also point out that left- and right-handed
fermions do not decouple if one regulates the theory using an
equal-time spatial circle (i.e., anti-periodic boundary conditions on
$x^1$) as used in \cite{les94}. In simple terms the argument is that
when one rolls the coordinate $x^1$ the lines $x^+=constant$ along
which the right-handed fermions propagate and the lines $x^-=constant$
along which the left-handed fermion propagate will intersect an
infinite number of times.   By contrast, in flat space and with a
circle in $x^-$ these line only intersect once.  Thus this method of
regularization is more natural.

The claims of Kutusov and Schwimmer have never been numerically
checked and we hope that this calculation is the first step in that
direction.  Checking these relations would present a particularly
interesting challenge for lattice gauge theory since these are
relations between theories with even and odd numbers of fermions.
 Among other interesting results we would expect to see is the
screening of ``fractional'' charges, e.g., charges in the fundamental
representation of $SU(N)$ when the dynamical fermions are in the
adjoint representation \cite{grk95}.  It has also been shown that when
the dynamical fermions are given a mass the screening is replaced by
true confinement, with a string tension given by
\begin{equation}
\sigma =2 m \Sigma\; .
\end{equation}
This relation also indirectly shows the deconfinement at zero mass \cite{grk95}.

\section{ Conclusions }

We have shown that in $QCD$ coupled to adjoint fermions in two dimensions the
light-front vacuum is two-fold degenerate as one would expect on general grounds. 
The source of this degeneracy is quite simple.  Because of the existence of Gribov
copies, the one gauge degree of freedom, the zero mode of $A^+$, must be
restricted to a FMD.  The domain of this variable, which after normalization we
call $Z_R$, is bounded by the integers.  Furthermore there is a $T_1R$ symmetry
which for $Z_R$ is simply a reflection about the midpoint of the FMD. The $T_1R$
symmetry operator acting on the fermionic Fock vacuum generates a second
degenerate state.  Since $T_1$ generates a spectral flow for the fermions this
second vacuum contains one left-handed and one right-handed fermion each with
momentum ${\pi/ 2L}$, but the state can be shown to still have $P^+=P^-=0$. We
form the analog of a $\theta$ vacuum from these degenerate vacuum states which
respects all of the symmetries of the theory. 

We found that the zero mode of $J_3^R$ includes a gauge correction $Z_R(x^+)$, and
since $Z_R(x^+)$ is dynamical it is impossible to satisfy the zero mode
of Gauss' law for the color three component. Furthermore this problem
reappears in the $\pm$ color components of Ampere's law.  We show that  this
condition on the zero mode of $J^3_R$ can hold weakly as a
condition on matrix elements between physical states in the large $L$ limit. 

We find that $Tr({\bar \Psi} \Psi)$ has a vacuum expectation value with respect to
this $\theta$ vacuum and we find an exact expression for this condensate $\Sigma$.
It is unlikely that the condensate we find here is equivalent to that found in
\cite{les94}, because the infrared regulator used there appears to couple the left
and right-handed fermions.

Since $P^-$ is block diagonal in our degenerate vacuum states the condensate does
not affect the massive spectrum of the theory and the theory has no massless bound
state. However this condensate is proportional to the string tension
characterizing the force between fractional charges when the adjoint fermions are
given a small mass.

\acknowledgments
\noindent
This work was supported in part by a grant from
the US Department of Energy. S.S.P would like to acknowledge the
hospitality of the Aspen Center for Physics where part of this work
was completed, David Kutasov for many helpful and instructive
conversations and Syd Meshkov for a careful reading of the manuscript.

\end{document}